# Highly Modulated Dual Semimetal and Semiconducting γ-GeSe with Strain Engineering

Changmeng Huan [1, 2], Pu Wang [1, 2], Binghan He [1, 2], Yongqing Cai [3, *], Qingqing Ke[1, 2, *]

[1] School of Microelectronics Science and Technology, Sun Yat-sen University, Zhuhai 519082, China

[2] Guangdong Provincial Key Laboratory of Optoelectronic Information Processing Chips and Systems, Sun Yat-sen University, Zhuhai 519082, China

[3] Joint Key Laboratory of the Ministry of Education, Institute of Applied Physics and Materials Engineering, University of Macau, Taipa, Macau, China

\* Corresponding authors

E-mail: yongqingcai@um.edu.mo; keqingq@mail.sysu.edu.cn

## Abstract

Layered hexagonal γ-GeSe, a new polymorph of GeSe synthesized recently, shows strikingly high electronic conductivity in its bulk form (even higher than graphite) while semiconducting in the case of monolayer (1L). In this work, by using first-principles calculations, we demonstrate that, different from its orthorhombic phases of GeSe, the γ-GeSe shows a small spatial anisotropic dependence and a strikingly thickness-dependent behavior with transition from semimetal (bulk, 0.04 eV) to semiconductor (1L, 0.99 eV), and this dual conducting characteristic realized simply with thickness control in γ-GeSe has not been found in other 2D materials before. The lacking of d-orbital allows charge carrier with small effective mass (0.16 $m_0$ for electron and 0.23 $m_0$ for hole) which is comparable to phosphorene. Meanwhile, 1L γ-GeSe shows a superior flexibility with Young's modulus of 86.59 N/m, only one-quarter of that of graphene and three-quarters of that of $MoS_2$, and Poisson's ratio of 0.26, suggesting a highly flexible lattice. Interestingly, 1L γ-GeSe shows an in-plane isotropic elastic modulus inherent with hexagonal symmetry while an anisotropic in-plane effective mass owing to shifted valleys around the band edges. We demonstrate the feasibility of strain engineering in inducing indirect-direct and semiconductor-metal transitions resulting from competing bands at the band edges. Our work shows that the free 1L γ-GeSe shows a strong light absorption (~$10^6$ $cm^{-1}$) and an indirect bandgap with rich valleys at band edges, enabling high carrier concentration and a low rate of direct electron-hole recombination which would be promising for nanoelectronics and solar cell applications.

**Keywords:** γ-GeSe, Isotropic elastic properties, Strain effect, Indirect-direct bandgap transition

# 1. Introduction

Group IV−VI monochalcogenides (GeS, GeSe, SnS, SnSe, etc), an emerging two-dimensional (2D) layered system with the puckered structure similar to phosphorene, have attracted enormous interest due to their potential in electrocatalytic, nanoelectronics, optoelectronic, thermoelectric and piezoelectric applications[1-7]. Germanium selenide (GeSe), with highly anisotropic electronic and optical properties due to the puckered structure, has aroused extensive attention for optoelectronics owing to its excellent stability and environmental sustainability than other Ⅳ−Ⅵ monochalcogenides[8-13]. For instance, the anisotropic optoelectronic properties of GeSe have been designed for high-performance polarization-sensitive photodetector exhibiting a low dark current of approximately 1.5 pA, a quick response of 14 μs, and a high detectivity of $4.7 \times 10^{12}$ Jones[14]. Furthermore, the rippling ferroic phase transition and domain switching have been realized in monolayer GeSe owing to its unique puckered structure[15].

However, the above-mentioned applications of GeSe are all based on the conventional orthorhombic phase. First-principles calculations proved that a new hexagonal phase of GeSe (named as γ-GeSe) can be stabilized in monolayer limit[16]. More recently, Lee et al. synthesized and identified the hexagonal phase of GeSe with two merged buckled honeycomb lattices for the first time[17]. The bulk form of this new polymorph of GeSe shows a surprisingly higher electrical conductivity of $3\times10^5$ S/m than graphite. Kim et al. verified a reversible and strain-tunable spontaneous polarization and spin-splitting in few-layer and bulk γ-GeSe through first-principles calculations, which makes γ-GeSe promising in nanoelectronics and warrants further studies on its physical properties in 2D form[18]. The layered γ-GeSe is liable to be mechanically exfoliated into a strain-sensitive monolayer (1L). Strain engineering is considered an efficient route to modulate the electronic structures of 2D materials[19-22]. The strain effect on electronic structures can modify the bandgap size, change the effective mass of carriers, and induce direct/indirect bandgap transitions, which in turn affects the electronic and optical properties[23-25]. Unfortunately, the strain effect on modulation of the electronic structure for γ-GeSe is still unknown albeit critically important for nanoelectronics and chemical applications.

In this work, we explore the strain- and thickness-dependent electronic properties of this novel phase of GeSe via first-principles calculations. Our results indicate that the valence and conduction band edges mainly consist of hybridized state of s-/p-orbitals of Ge and Se with small effective mass comparable to phosphorene. Different from its orthorhombic phases, the hexagonal GeSe shows a strong thickness modulated transition of semimetal (bulk) to semiconductor (1L), which allows a strong modulation of electronic properties via thickness engineering. The hexagonal GeSe shows nearly isotropic elastic properties and a small Young's modulus (86.59 N/m), suggesting a highly flexible nature for wearable and flexible electronic devices. The strain effect on the bandgaps and effective masses shows a strong non-linear behavior owing to competing bands at the band edges. Our work shows that the 1L γ- GeSe has an indirect band gap with rich distribution of valleys at band edges, implying high carrier concentration and suppressed electron-hole recombination thus highly appealing for nanoelectronics and solar cell.

## 2. Methods

The first-principles calculations were performed on the projected augmented wave

(PAW) pseudopotentials as implemented in Vienna ab initio simulation package (VASP) [26, 27]. The electron exchange-correlation interactions were treated with the generalized gradient approximation (GGA) using the Perdew-Burke-Ernzerhof (PBE) functional. The semiempirical vdW-D2 method was adopted to better describe the van der Waals (vdW) interaction[28]. The convergence thresholds for electronic and ionic relaxations were set to $1.0 \times 10^{-6}$ eV and 0.005 eV Å$^{-1}$, respectively. The Brillouin zones were integrated using the Gamma-centered k-points sampling with a reciprocal space resolution higher than $2\pi \times 0.02$ Å$^{-1}$ for geometry optimization and electronic structure calculations. Since the PBE method severely underestimates the band gap, the electronic structure was also performed using the standard Heyd-Scuseria-Ernzerhof (HSE06) hybrid functional[29].

## 3. Results and discussion

### 3.1. Geometry structure of 1L γ-GeSe

The bulk γ-GeSe (space group *P6₃mc*) is in a dynamically stable A-B' stacking arrangement (figure 1(a)), where Ge and Se atoms are bonded to form a puckered 2D honeycomb structure. The optimized lattice parameters are $a = b = 3.77$ Å and $c = 15.39$ Å, agreeing well with the experimentally reported results ($a = b = 3.73$ Å, $c = 15.40$ Å) [17]. In addition, as shown in Table S1 (refer to supplementary information), the total energies ($E_0$) and formation energies ($E_f$) of 1L γ-GeSe are 0.20 and 0.21 eV lower than these of orthorhombic phase (α-GeSe), indicating that it is more stable. Meaningfully, the exfoliation energy ($E_{exf}$) of γ-GeSe is 16.03 meV/Å$^2$ lower than that of α-GeSe, which make it liable to be exfoliated into a monolayer. Therefore, the study of 1L γ-GeSe has a far-reaching practical significance.

To build 1L γ-GeSe unit cell (figure 1(b)), the bulk unit cell is first cleaved along the (001) plane, then a 20 Å vacuum layer is inserted along the [001] orientation. After structural relaxation, the obtained lattice constant for the unit cell is $a = b = 3.76$ Å. The 1L (1 × $\sqrt{3}$) supercell (figure 1(c)) is designed to investigate the elastic properties and the effective masses along the zigzag and armchair directions. The Brillouin zones of the unit cell and supercell are plotted in figure 1(d), where the Γ-X and Γ-Y path in $k$ space are corresponding to the zigzag and armchair directions marked in figure 1(c), respectively.

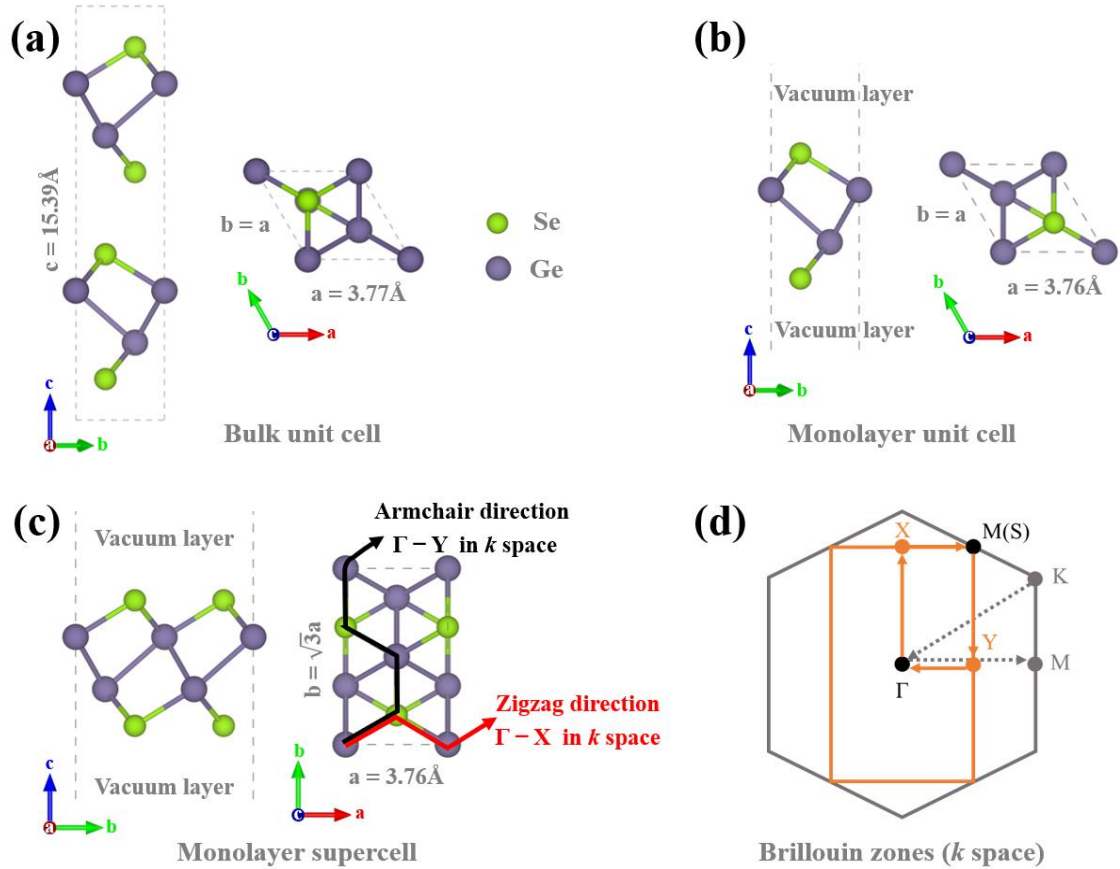

**Figure 1.** Front and top views of the bulk γ-GeSe unit cell (a), unit cell (b) and 1× $\sqrt{3}$

supercell (c) of 1L γ-GeSe. (d) The Brillouin zones (*k* space) of the unit cell (grey hexagon) and the 1× $\sqrt{3}$ supercell (orange rectangle) of 1L γ-GeSe.

**3.2. Electronic structure of 1L and bulk γ-GeSe**

The band structures of bulk γ-GeSe calculated by PBE and HSE are shown in figure 2(a) and (b), respectively. The results in PBE method show that there is an overlap (0.17 eV) between the conduction band minimum (CBM) and the valence band maximum (VBM) of bulk γ-GeSe, exhibiting a semi-metallic nature, which is consistent with the previous result[17]. In HSE06 method, there is an extremely narrow bandgap of 0.04 eV with the VBM at Γ point and the CBM at $M_0$ point (0.43, 0, 0) between the Γ and M points. This phenomenon also occurs in $Ag_2Se$ and $TiS_2$, where the band-gap openings only take place with the HSE06 function, as there is no bandgap if the PBE functional is used instead[30, 31]. Such a tiny gap in bulk γ-GeSe however is comparable with and easily overcome by the thermal excitation (0.026 eV) at room temperature. Therefore, a substantial thermal-induced conduction of carriers could be expected in bulk γ-GeSe and much higher than its cousin of bulk α-GeSe which has a substantial band gap of 1.24 eV (HSE06 value)[12]. Such a quasi-metallic behavior of γ-GeSe is highly appealing for lowing the interfacial contacting resistance in nanoelectronics devices.

It is highly interesting to examine the quantum confinement effect of γ-GeSe to see if the semi-metallic behavior is retained in atomically thin limit. For 1L γ-GeSe, the bandgap is opened up to 0.60 and 0.99 eV for PBE and HSE06 method, respectively, with the CBM shift to Γ point and the VBM shift to $K_1$ point (0.05, 0.05, 0) between Γ and K points. The direct bandgaps at Γ point are 1.02 and 1.32 eV for PBE and HSE06 method, respectively, close to the previous theoretical prediction for the A-A and A-B stacked γ-GeSe[16]. It is worth noting that there are satellite states around the M point (0.5, 0, 0), which are almost degenerate with the CBM and are ~41 meV higher than the CBM at the Γ point. Similarly, there is a hump-like structure in the top valence band near the Γ point, where the $M_1$ point is 34 meV lower than the $K_1$ point. The nearly degenerate energy valleys potentially allow the segregation of electrons and/or holes of the same energy by momentum and for valley effects to manifest. The rich distribution of nearly degenerate valleys around the band edges potentially allows the segregation of electrons and/or holes of the same energy by momentum and for valley effects to manifest, which are potential applications for valleytronics and solar cells[32]. Importantly, a broadening of band gap up to nearly 1 eV with bulk/1L phases of 0.04/0.99 eV (HSE06 value), rooted in the quantum confinement effect[33, 34], corresponds to a nontrivial transition from semi-metal for bulk to semiconductor for 1L γ-GeSe. The semi-metal and semiconducting dual roles in a single material realized simply with thickness control has not been found in other 2D materials before.

The bilayer (2L) γ-GeSe was also calculated for comparison. For the 2L γ-GeSe, the bandgap is 0.34 and 0.62 eV for PBE and HSE06 method (figure S1, refer to supplementary information), respectively. The CBM is located at $M_0$ point (0.43, 0, 0) consistent with the bulk, while the VBM is shifted to the $K_1'$ point (0.04, 0.04, 0) between the Γ and K points. The topology of the band structure for 2L γ-GeSe is similar to that of the 1L case in figure 2(c) and (d).

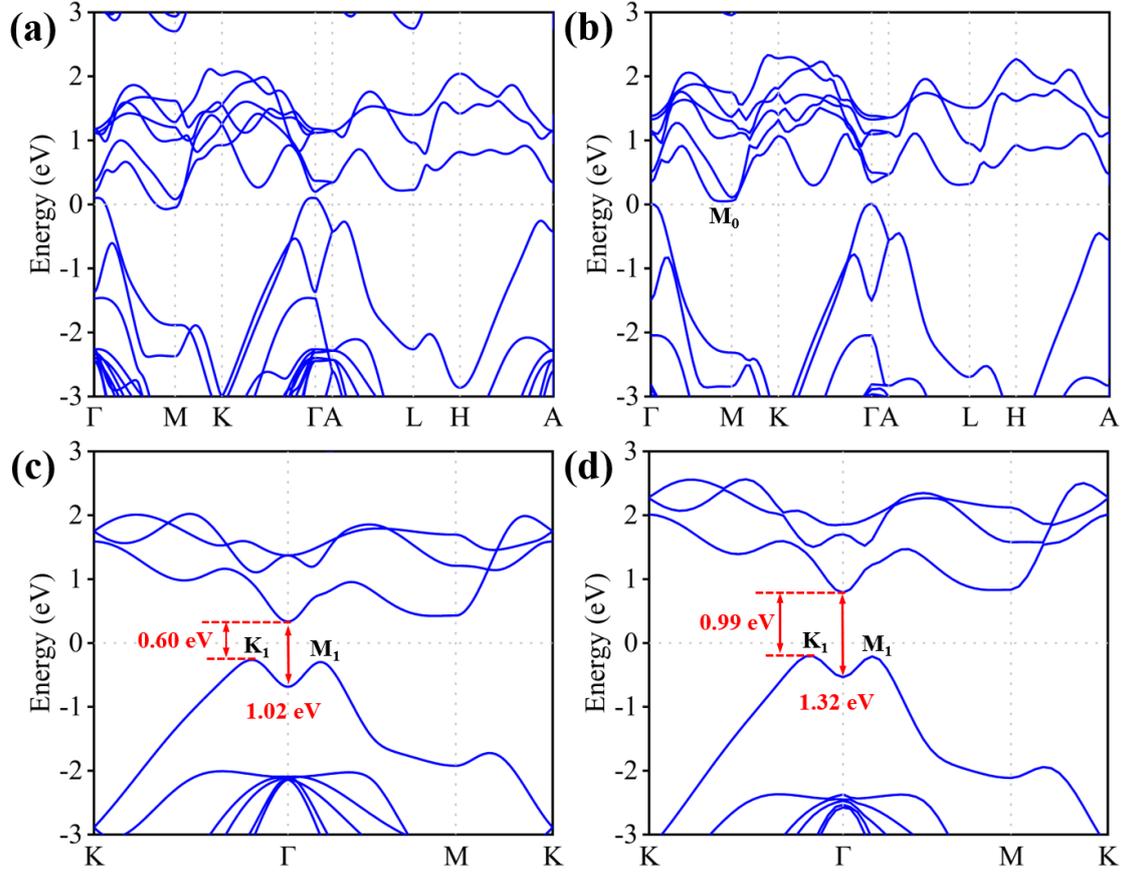

**Figure 2.** Band structures of bulk γ-GeSe (a and b) and 1L γ-GeSe (c and d) calculated by PBE (left panels) and HSE06 (right panels) method, indicating a transition of semimetal to semiconductor from bulk to 1L γ-GeSe.

With the aim to an in-depth analysis of the electronic structure, the projected band structures of 1L γ-GeSe were calculated in PBE method and presented in figure 3. The projected band structures of Ge (figure 3(a)) and Se (figure 3(a)) indicate that both Ge and Se atoms contribute to the conduction bands and the valence bands. The VBM mainly consists of Ge 4s and Se $4p_z$ orbitals, while the CBM is composed of Ge 4s, Ge 4p$_z$, and Se 4s orbitals with a little Se 4p$_z$ components. The isosurfaces of wavefunctions in real space of the CBM and VBM are plotted in figure 3(c) to further improve our understanding. In the CBM case, the wavefunctions consist of blue and yellow irregular ellipsoids, which is formed by superposition of spherical 4s and dumbbell-shaped 4p$_z$ orbital of Se and Ge in the crystal field, respectively. In the VBM case, the wavefunctions are composed of blue and yellow irregular spherical Ge 4s orbitals and irregular dumbbell-shaped Se 4p$_z$ orbitals in the crystal field. The partial charge densities of the CBM and VBM in figure 3(d) show that the VBM tend to be bonding-like, while the CBM is nonbonding-like.

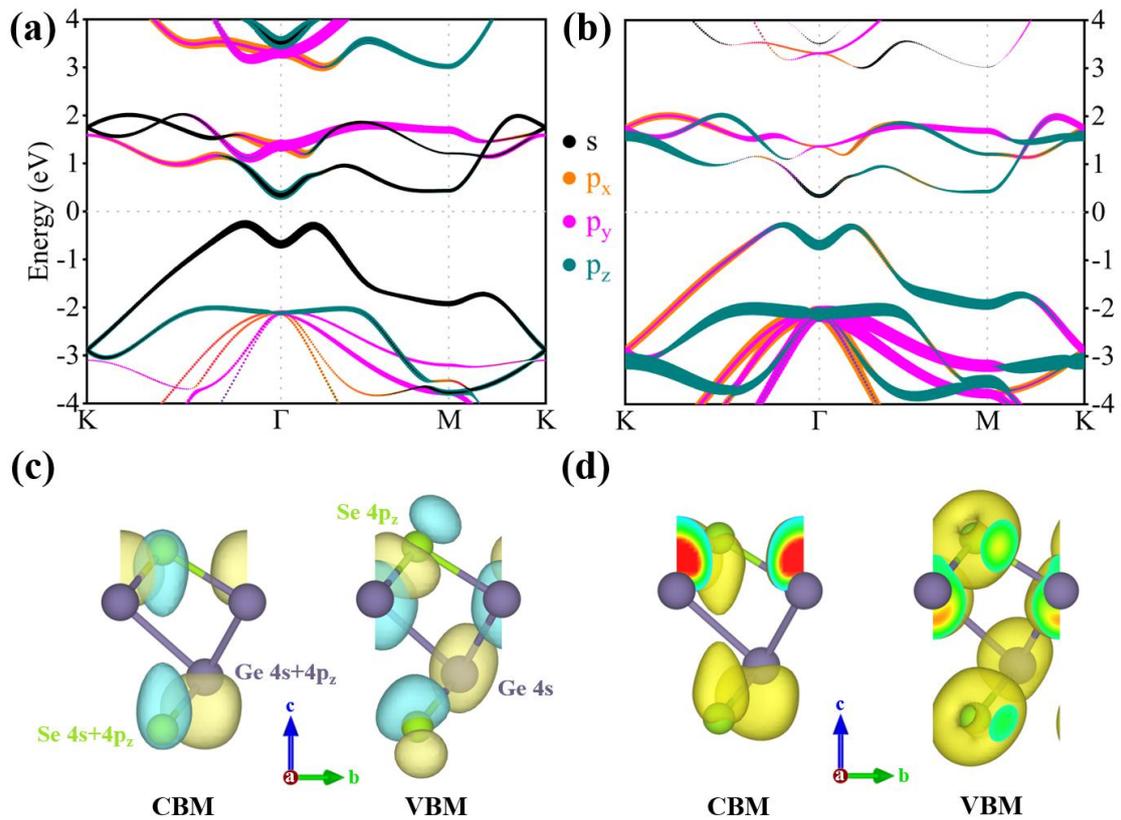

**Figure 3.** Projected electronic band structures of different orbitals of Ge (a) and Se (b) in 1L γ-GeSe. The size of the solid circles represents the weight of the orbital

contribution to the band. (c) The isosurfaces of wavefunctions in real space for the CBM at the Γ point and the VBM at the $K_1$ point, respectively. (d) Partial charge densities of the CBM and VBM states (isosurface is set to 0.005 e/Å$^3$).

The projected density of states (PDOS) in figure 4(a) shows that the total DOS of γ-GeSe is composed of five discrete zones within a broad range between -14 and 6 eV. The number of electrons in each zone was calculated through

$$n = \int_{E_0}^{E_1} D(E) \, dE \tag{1}$$

where $n$ represents the number of electrons, $D(E)$ represents the total DOS, $E_0$ and $E_1$ represent the upper limit and lower limit of the energy of the desired band, respectively. The results show that the total number of electrons for unit cell (Ge$_2$Se$_2$) is 20, including 4 electrons for zone-1, 4 electrons for zone-2, and 12 electrons for zone-3. Zone-4 and -5 above Fermi level contain 6 empty electronic states. Given that Ge and Se elements occupy the same row and are close neighbors in periodic table, the differences in electronic configurations and electronegativity between them is small, which leads to complex orbital interactions[35], and various hybridizing possibilities. In order to quantitatively describe the component of atomic orbitals within each zone, the integrated PDOS was utilized to disentangle the ratio of a specific orbital subshell as

$$\eta = \frac{\int D_{n,m}(E) \, dE}{\int D_n(E) \, dE} \tag{2}$$

where $\eta$ represents the proportion of $m$ orbital to the $n$th band, $D_{n,m}(E)$ is the PDOS of $m$ orbital in the $n$th band, and $D_n(E)$ is the total DOS of all orbitals in the $n$th band. The decomposed results are summarized in figure 4(b), where the Ge 4s/4p-orbitals and Se 4s/4p-orbitals hybridize and couple, giving rise to three filled orbital regions and two unoccupied orbital regions. It is found that the Se 4s orbitals contribute 90% of the zone-1 (the deepest band at around -14 eV), resulting in σ-bonding orbitals. The composition of zone-2 between -10 and -6 eV is mainly Ge 4s (70%) and Se 4s (13%), while a small amount of s-p coupling from Se 4p (14%) being involved. Zone-3 between -6 and 0 eV is composed of Se 4p (66%), Ge 4s (7%) and Ge 4p (26%), including both π (p-p coupling) and σ (mainly s-p coupling) bonding characteristics. More interestingly, the contribution of Ge 4s (7%) is mainly distributed at the VBM (figure 3(a)), which may be due to the asymmetrically layered-crystal structure arising from the stereo-chemically active lone pairs[36, 37]. Zone-4 acts as the conduction band and consists of Ge 4p (42%) and Se 4p (42%), as well as a slight s-p coupling from Ge 4s (12%). The zone-5 is mainly contributed by unoccupied Ge 4p orbitals and partly by unoccupied 4p orbitals of Se, exhibiting π* antibonding orbitals of p-p coupling.

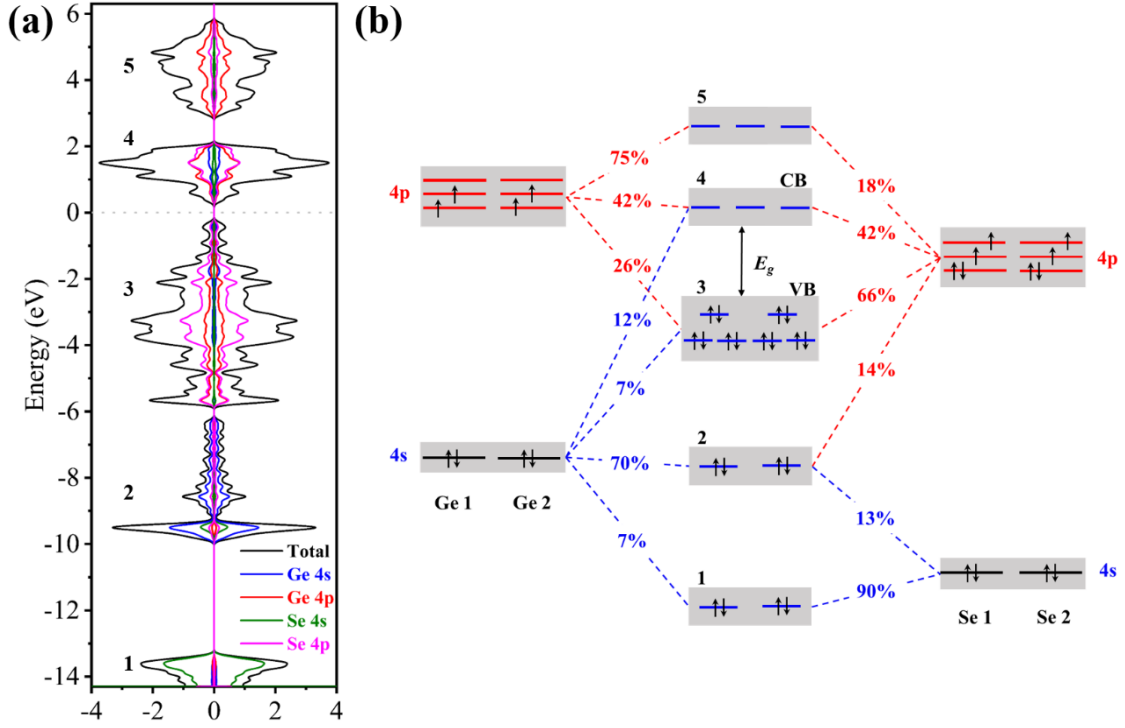

**Figure 4.** (a) The PDOS of 1L γ-GeSe calculated in PBE method. (b) Schematic diagram of the averaged weight of atomic 4s and 4p orbitals of Ge and Se in five zones of PDOS from -14 to 6 eV for 1L γ-GeSe. Only the orbitals with weight greater than 5% are listed.

### 3.3. Highly flexible and nearly isotropic elastic properties

In order to explore the elastic properties and the strain effect on the effective mass along the zigzag and armchair directions, the orthogonalized $(1 \times \sqrt{3})a$ supercell was used in the calculation. In the orthogonalized cell, the relation between strain and stress can be expressed as[38]

$$\begin{pmatrix} \sigma_1 \\ \sigma_2 \\ \sigma_6 \end{pmatrix} = \begin{pmatrix} C_{11} & C_{12} & 0 \\ C_{21} & C_{22} & 0 \\ 0 & 0 & C_{66} \end{pmatrix} \begin{pmatrix} \varepsilon_1 \\ \varepsilon_2 \\ \varepsilon_6 \end{pmatrix} \quad (3)$$

where $C_{ij}$, $\sigma_i$, and $\varepsilon_i$ represent the elastic stiffness tensor, stress, and strain with three independent components ($i,j$ = 1,2,6), respectively. The subscript 1,2, and 6 corresponds to xx, yy and xy directions, respectively. Then, the elastic strain energy per unit area based on the strain-energy method can be written as[39]

$$\frac{\Delta E (S, \varepsilon_i)}{S_0} = \frac{1}{2} ( C_{11}\varepsilon_1^2 + C_{22}\varepsilon_2^2 + 2C_{12}\varepsilon_1\varepsilon_2 + C_{66}\varepsilon_6^2 ) \quad (4)$$

where $S_0$ and $\Delta E$ are the equilibrium area of 1L γ-GeSe and the elastic strain energy, respectively. Therefore, the $C_{ij}$ can be expressed as the second-order partial derivative of $\Delta E$ to $\varepsilon_i$, that is, $C_{ij} = (1/S_0)(\partial^2\Delta E/\partial\varepsilon_i \partial\varepsilon_j)$ in unit of the force per unit length (N/m or GPa·nm). Our calculated results of γ-GeSe, compared with values of some other mainstream 2D materials and reported values from literature are summarized in Table 1. As shown in the table, the $C_{11}$ of γ-GeSe is much lower than those of $MoS_2$ and graphene, and comparable to phosphorene, indicating a relatively soft nature of the lattice.

**Table 1**

PBE-calculated in-plane elastic stiffness tensor (in units of N/m) of γ-GeSe, graphene, $MoS_2$ and phosphorene, as well as the values reported in the literature.

| | this work | | | | literature | | | |
|---|---|---|---|---|---|---|---|---|
| | $C_{11}$ | $C_{12}$ | $C_{22}$ | $C_{66}$ | $C_{11}$ | $C_{12}$ | $C_{22}$ | $C_{66}$ |
| Graphene | 354.3 | 62.2 | / | / | 358.1[40] | 60.4[40] | / | / |
| $MoS_2$ | 133.5 | 35.1 | / | / | 131.4[41] | 32.6[41] | / | / |
| Phosphorene | 107.6 | 21.7 | 33.8 | 27.9 | 105.2[42] | 18.4[42] | 26.2[42] | / |
| **GeSe** | **93.0** | **24.3** | **92.9** | **34.2** | / | / | / | / |

The Young's modulus, shear modulus and Poisson's ratio of 1L γ-GeSe were calculated and presented in figure 5. For comparison, the elastic properties of phosphorene were also calculated. As shown in figure 5(a-c), the Young's modulus, shear modulus and Poisson's ratio of phosphorene all show strong anisotropy, while 1L γ-GeSe shows nearly isotropic elastic properties. The Young's modulus and Poisson's ratio of phosphorene along the zigzag and armchair direction are 93.65 and 29.44 N/m, and 0.64 and 0.20, respectively, which are in good agreement with the previous results[25]. Surprisingly, the Young's modulus and Poisson's ratio of 1L γ-GeSe along the zigzag and armchair direction are 86.61 and 86.53 N/m, and 0.26 and 0.26, respectively. These nearly isotropic elastic properties of 1L γ-GeSe in different directions can be clearly seen in figure S2 (refer to supplementary information). To compare with graphene and $MoS_2$ in the literature, where the elastic properties were calculated in the form of unit cell, we also calculated the elastic properties of 1L γ-GeSe, graphene, and $MoS_2$ in unit cell. As shown in figure S3 (refer to supplementary information), the calculated Young's modulus of 1L γ-GeSe (86.59 N/m) is much smaller than that of 1L $MoS_2$ (124.28 N/m) and graphene (343.33 N/m), which is consistent with the values in the literature[43, 44]. The above results suggest that 1L γ-GeSe is highly flexible and may have potential applications in wearable and flexible electronic devices.

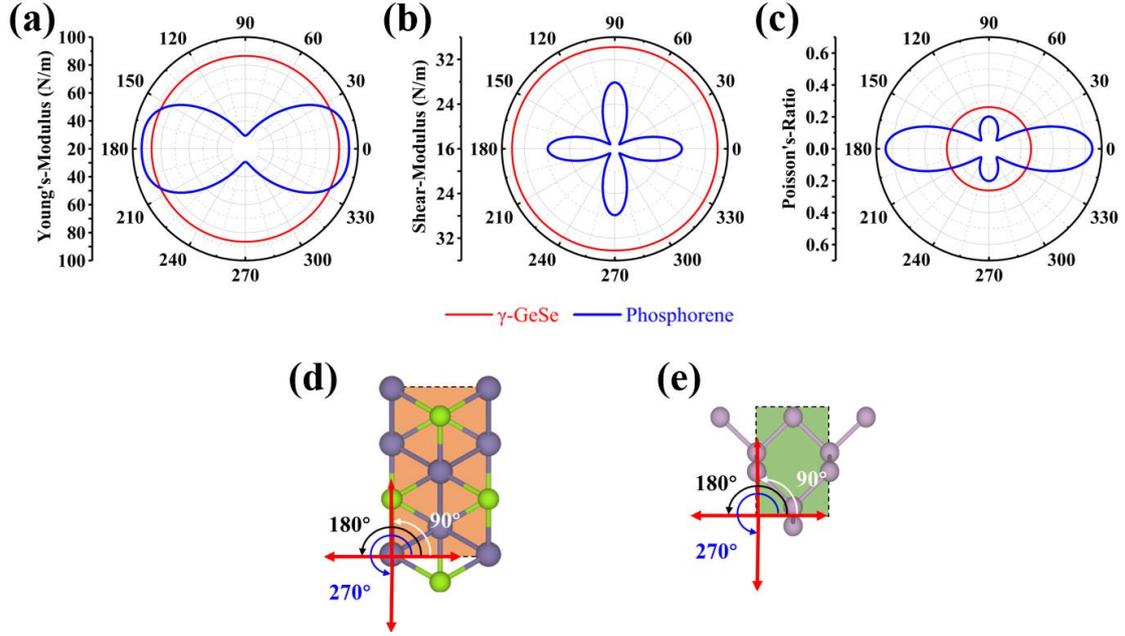

**Figure 5.** The Young's modulus (a), shear modulus (b), and Poisson's ratio (c) of 1L γ-GeSe and phosphorene calculated in PBE method. Top views of the geometric structure of 1L γ-GeSe supercell (d) and phosphorene (e).

### 3.4. Strain tunable indirect-direct bandgap transition

We next explore the effect of strains on the electronic band structure of 1L γ-GeSe. Since previous studies suggested that the trends of strain-dependent band structure and bandgap calculated by PBE method are consistent with the behaviors by HSE06 method[25], here only the PBE method is adopted to compare the evolution of band structures under biaxial strains. It is found that strains ($\varepsilon$) have significant effect on the band structure especially for branches along Γ to X (0.5, 0, 0) and Γ to Y (0, 0.5, 0) directions, which correspond to the zigzag and armchair directions (figure 1(c)), respectively. For simplification, only the bands along these two directions were plotted in figure 6. Interestingly, while the 1L γ-GeSe without strain has an indirect bandgap with the CBM (VBM) at the Γ (B) point and this indirect nature maintains under compressive strain, there is an indirect-direct transition at +14% tensile strain. The underlying mechanism of this indirect-direct transition is the relative alignment of the valleys of the frontier orbitals: the A and B maximums at the VBM and the C and D minimums at the CBM. With exerting a moderate strain up to +4%, the level of D valley as the CBM drops slightly. With further increasing tensile strain, the level of D state undergoes a strong upward shift which makes the C valley becoming the CBM at $\varepsilon$ = +14%. Similar competing bands near the band-edges are also found in valence band. Under tensile strains, the original VBM at the B point moves downward and becomes lower in energy than that of A. Thus, the 1L γ-GeSe transitions from indirect to direct bandgap due to the band exchanging and distortion under tensile strains. In the compressive strain considered, the VBM is always located at state B, while the level of C valley reduces dramatically and the CBM moves from D to C valley.

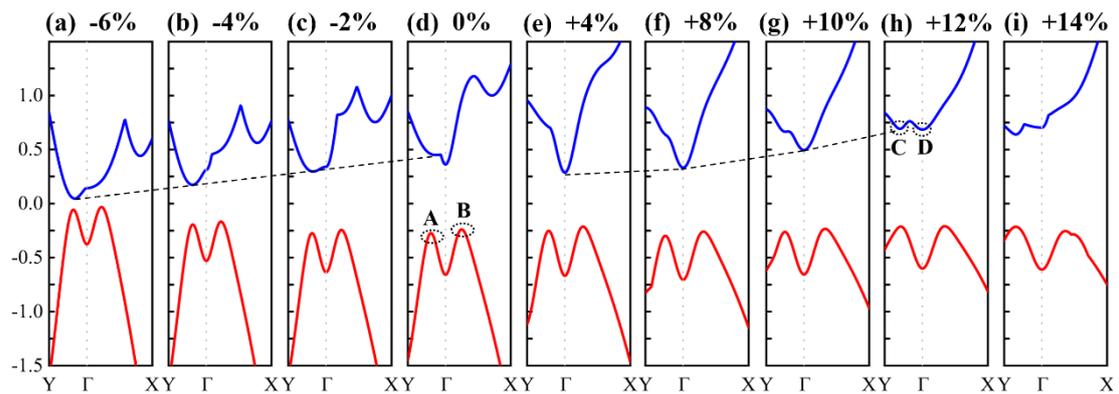

**Figure 6.** Strain-tunable indirect-direct bandgap transition in 1L γ-GeSe by PBE method. Positive values represent tensile strains, while negative values correspond to compressive strains. The Fermi energy is set as zero and the dashed lines are guide for eye only to show the energy shifts of states C and D.

Figure 6 demonstrates that the shift of band edges and bandgaps are highly sensitive to strain. The quantitative relationship between the bandgap *vs* strain was plotted in figure 7. In terms of compressive strain, the bandgap keeps decreasing due to the downward shift of the CBM, which originates from the strong shift of the C valley. At $\varepsilon = -6\%$, the bandgap drops sharply to 0.08 eV from 0.34 eV at $\varepsilon = -4\%$ strain and finally disappears at $\varepsilon = -8\%$. The decreasing bandgap is related to the variation of bond length as the compressive strain increases. As shown in figure S4 (refer to supplementary information), Ge-Se bond lengths projected in-plane (a/b direction) shorten while the out-of-plane components (c-direction) become longer under compressive strain, which results in a decrease of the $4p_z$ components in the valence band (shaded area) and the CBM, and accordingly an increase of $4p_y$ and $4p_x$ components near the CBM (figure S4(c)). The projected band structures (figure S5, refer to supplementary information) show that the top valence band extends upwards, while the conduction band at the red arrow moves downwards, which is associated with the reduction of bandgaps. As shown in figure S6 and S7 (refer to supplementary information), the effect of tensile strain (increase) is in line with compressive strain (decrease). The bottom conduction band first moves down to reduce the bandgap for strains (0-4%), and then becomes flat and moves up to increase the band gap for strains (6%-12%). Therefore, the bandgap first decreases from 0.60 ($\varepsilon = 0$) to 0.50 eV ($\varepsilon = 4\%$), and then rises with further expansion (6%-12%). The bandgap reaches to a maximal value of 0.89 eV at $\varepsilon = 12\%$ and the CBM switches from state D to state C thereafter. Our work also reveals that, around the equilibrium structure and under moderate strains, the electronic band gap is overall more sensitive to the compressive strain than the same amount of tensile strain.

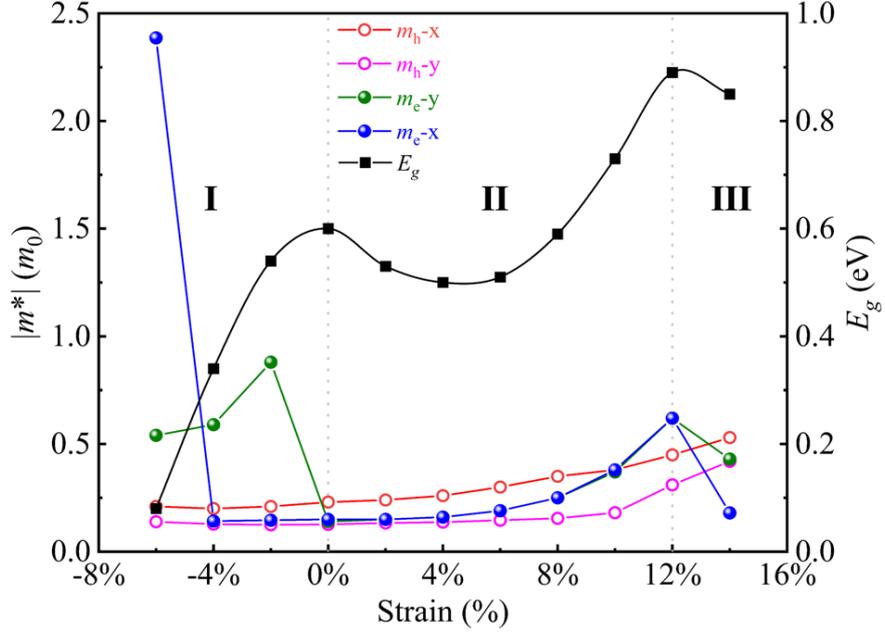

**Figure 7.** The bandgaps and effective masses of electrons $m_e$ and holes $m_h$ (in units of free electron mass $m_0$) along different directions as a function of biaxial strains in 1L γ-GeSe obtained by PBE calculations. Positive (negative) strain values represent tensile (compressive) strains.

The effective masses of the electron and hole can be obtained from the band structure according to the formula

$$m^* = \hbar^2 \left(\frac{d^2E}{dk^2}\right)^{-1} \tag{5}$$

For 1L γ-GeSe without strain, the effective masses of electrons ($m_e$) are 0.15 and 0.14

$m_0$ along zigzag and armchair directions, respectively. The effective masses of holes ($m_h$) are 0.23 and 0.13 $m_0$ along zigzag and armchair directions, respectively. Both masses are small and comparable to phosphorene which is well-known for its light electron/hole carriers, indicating that 1L γ-GeSe tends to possess unique performance of carrier transport.

The effective masses as a function biaxial strain were calculated and plotted in figure 7. Similar to the situations of the band structure, the effective masses evolve differently according to the three different strain zones: zone-I (compressive strains from -6% to 0), zone-II (tensile strains from 0 to 12%), and zone-III (tensile strains beyond 12%). According to the Formula (5), effective mass is inversely proportional to the curvature of the valleys (peaks) at the conduction (valence) band. As shown in figure 6, the top valence band contracts with the increase (decrease) tensile (compressive) strain, resulting in a decrease in curvature at the peak and thus an increase in effective mass. Therefore, $m_h$ slightly increases with the expanding of lattice in the three zones. For electrons in zone-I, the CBM shifts to the C valley where its curvature increases with the strains, resulting in the effective mass decreases from 0.88 $m_0$ ($\varepsilon = -2\%$) to 0.54 $m_0$ ($\varepsilon = -6\%$) along the armchair direction. In the zigzag direction (Γ-X), the effective mass depends on the curvature near the Γ point, which initially remains unchanged and then decreases abruptly at $\varepsilon = -6\%$, resulting in a sudden increase in effective mass to 2.39 $m_0$. In zone-II, the CBM remains at the Γ point, its curvature first remains constant (0-4%) and then decreases with the increase of strains (6%-12%), resulting in an increase in effective mass (6%-12%) and reaching the maximum value of 0.62 $m_0$ at $\varepsilon = 12\%$. As the tensile strain increase to zone-III, the CBM shifts to the C valley and the curvature near the Γ point increases, resulting in a decrease of effective masses along the armchair and zigzag directions.

**3.5. Strain effect on the optical absorption**

Given the inherent high flexibility of 1L γ-GeSe, it is of great significance to understand the strain effects on its optical properties. Therefore, the absorption spectra of 1L γ-GeSe under different strains were calculated and shown in figure 8. It is found that 1L γ-GeSe shows anisotropic optical absorption along the armchair and zigzag directions, with strong absorption in the ultraviolet (UV) and visible ranges, and a relatively weak absorption in the near-infrared (NIR). In NIR regions near the absorption edge, absorption edge exhibits a redshift with the decrease (increase) of compressive (tensile) strain. A similar red-shift occurs in the UV range, resulting in strong violet-blue absorption under tensile strain. In the visible range, there are complex peaks in the strain-dependent absorption spectra due to the electronic transitions at different K-points in the reciprocal space. Significantly, the equivalent absorption coefficient ($\alpha_a$, $\alpha_a = A/t$) of 1L γ-GeSe with thickness ($t$) of 4.18~4.89 Å and absorbance ($A$) of 5~15% in the optimized solar irradiation range 1.5-2.5 eV is on the order of $10^6$ cm$^{-1}$, which is in the same order as that of 1L $MoS_2$ and graphene[45, 46]. The strong absorption capability makes 1L γ-GeSe a promising candidate for solar energy and optoelectronic applications.

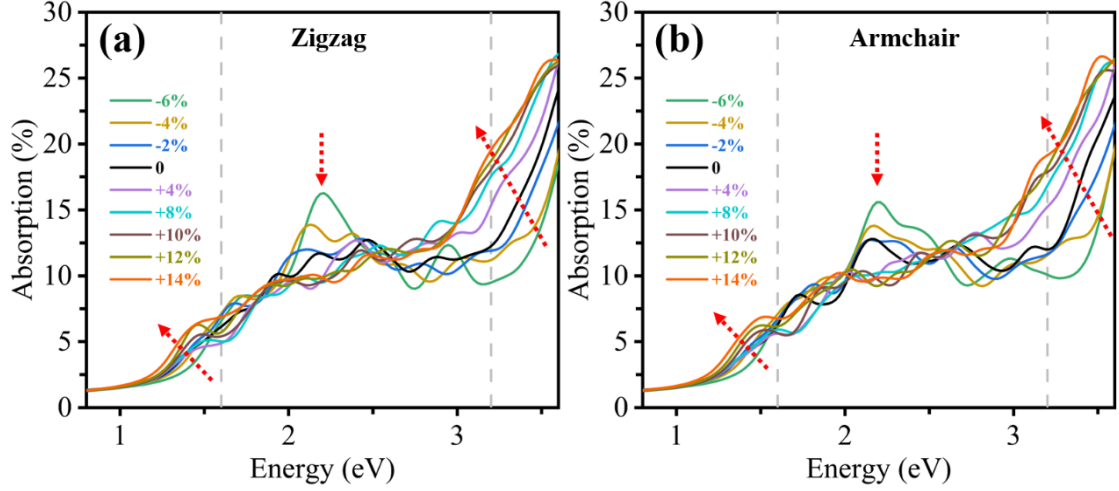

**Figure 8.** The absorption spectra of 1L γ-GeSe along the zigzag (a) and armchair (b) directions in HSE06 method.

Here we focus on the monolayer γ-GeSe. We would like to discuss on the potential aspect for obtaining the monolayer material. While the pioneering work demonstrated the viable synthesis of bulk γ-GeSe, no results of monolayer γ-GeSe was presented[17]. Nevertheless, it is possible to grow the monolayer material through tuning the growth parameters and conditions since in that work the bulk γ-GeSe has been found to be grown above vdW systems like h-BN. Other thinning methods like chemical exfoliation and mechanical exfoliation, which have been proven to be viable for obtaining monolayer dichalcognides, should also work for γ-GeSe. Moreover, we have recently shown the evidence of a fast lithiation within GeSe[47], suggesting an exfoliation via lithium intercalation for exfoliation.

## 4. Conclusions

In this work, we investigated the recently synthesized new phase of hexagonal GeSe through systematic calculations based on first-principles method. We found that the band edges of VB and CB consist of mainly hybridized state of s-/p-orbitals of Ge and Se, the VBM mainly contributed by Ge 4s and Se $4p_z$ orbitals and the CBM is composed of Ge 4s, Ge $4p_z$, and Se 4s orbitals with a little $4p_z$ components, which leads to small effective mass comparable to phosphorene. Different from its orthorhombic cousins, the hexagonal GeSe shows small anisotropic electronic property and a strong thickness-modulated transition of semimetal (bulk) to semiconductor (1L) which allows a strong modulation of electronic properties via thickness engineering. The 1L γ-GeSe shows nearly isotropic elastic properties. Moreover, the Young's modulus of γ-GeSe is much smaller than that of $MoS_2$ and graphene, suggesting a highly flexible nature for wearable and flexible electronic devices. We also calculated the strain effect on the electronic bandgap and effective mass which show a strong non-linear behavior owing to competing bands at the band edges. Our work indicates that 1L γ-GeSe has a strong light absorption (~$10^6$ $cm^{-1}$) and an indirect bandgap with rich distribution of valleys at band edges, implying a high carrier concentration and a suppressed rate of electron-hole recombination which are highly appealing for nanoelectronics and solar cell.

Finally, we would like to highlight the intriguing different isotropic elastic property (i. e. elastic modulus) while an anisotropic electronic property (such as effective mass) in γ-GeSe. The former is more or less related on the symmetry of the lattice in real space whereas the electronic property is dominated by the electronic momentum space. The γ-GeSe adopts a relatively high symmetry with in-plane hexagonal lattice similar to graphene, which induces an isotropic response to external stimuli. In contrast, the electron/hole acts according to Hamiltonian in the momentum space, and the presence of asymmetric pockets states around the band edge of γ-GeSe gives rise to asymmetric effective mass tensors and anisotropic behaviors.

## Competing financial interests statement

The authors declare no competing financial interests.

## Acknowledgments

The authors acknowledge the funding support from the 100 Talents Program of Sun Yat-sen University (Grant 76220-18841201), the National Natural Science Foundation of China (Grant 22022309) and Natural Science Foundation of Guangdong Province, China (2021A1515010024), the University of Macau (SRG2019-00179-IAPME, MYRG2020-00075-IAPME) and the Science and Technology Development Fund from Macau SAR (FDCT-0163/2019/A3).